\documentclass[a4paper,UKenglish,cleveref, autoref, thm-restate]{lipics-v2021}
\usepackage{threeparttable}

\title{Watts and Debts of Agentic Frameworks: An Empirical Study}

\author{Aneetta Sara Shany}{Software Engineering Research Center, IIIT Hyderabad, Hyderabad, India}{aneetta.sara@research.iiit.ac.in}{https://orcid.org/0009-0008-0130-0060}{}

\author{Chandrasekar Sridhar}{Software Engineering Research Center, IIIT Hyderabad, Hyderabad, India}{chandrasekar.s@research.iiit.ac.in}{https://orcid.org/0009-0009-3679-6331}{}

\author{Karthik Vaidhyanathan}{Software Engineering Research Center, IIIT Hyderabad, Hyderabad, India}{karthik.vaidhyanathan@iiit.ac.in}{https://orcid.org/0000-0003-2317-6175}{}

\authorrunning{Shany et al.} 

\Copyright{Aneetta Sara Shany, Chandrasekar Sridhar and Karthik Vaidhyanathan} 



\category{} 

\relatedversion{} 

\acknowledgements{This research was funded by the ANRF Prime Minister Early Career Research Grant (ANRF/ECRG/2024/003379/ENS).}

\nolinenumbers 

\EventLongTitle{ESEIW 2026 - 20th International Symposium on Empirical Software Engineering and Measurement (ESEM) - Registered Reports Track} 
\EventShortTitle{ESEM 2026}
\EventAcronym{ESEM}
\EventYear{2026}
\EventDate{October 4--9, 2026}
\EventLocation{Munich, Germany} 
\ArticleNo{32}

\begin{document}

\maketitle

\begin{abstract}

\emph{\textbf{Context.}} Every agentic AI system shipped to production carries two hidden risks: accumulated Technical Debt (TD) and unmonitored runtime energy costs. While functional benchmarking is common, the empirical link between internal structural quality (specifically TD) and dynamic energy consumption during execution remains unexplored, creating a blind spot for practitioners and organizations managing sustainability and operational budgets at scale.

\emph{\textbf{Goal.}} We propose a confirmatory empirical study correlating Self-Admitted Technical Debt (SATD) with hardware-level runtime energy consumption across agentic frameworks, to determine whether code quality can drive energy-aware design decisions.

\emph{\textbf{Method.}} We will evaluate five open-source agentic frameworks by executing a standardized task suite in a strictly controlled environment. SATD will be extracted via automated Python-based comment mining and categorized via LLM-based classification using a fine-tuned prompt, 
while runtime energy will be measured at the hardware level. Our study will investigate three core research questions: (RQ1) the presence of TD within these frameworks; (RQ2) the variance in runtime energy consumption across architectures; and (RQ3) the statistical correlation between a framework's TD and its task-level energy consumption.

\emph{\textbf{Conclusion.}} The findings will establish whether automated source code analysis can serve as a reliable, early-warning proxy for energy-efficient framework selection, thereby advancing both green software engineering and agentic AI quality research.

\end{abstract}

\section{Introduction}
\label{sec:intro}
Software engineering is undergoing a paradigm shift from single-prompt Large Language Model (LLM) interactions to autonomous, tool-using agentic frameworks. These frameworks have rapidly become the primary mechanisms for developing LLM-based production workflows~\cite{11316910}. Unlike simple query-response models, agentic systems execute in continuous, goal-directed loops comprising planning, memory retrieval, tool dispatch, self-reflection, and retry mechanisms~\cite{wang2024survey}, often running autonomously for minutes or hours.
Consequently, the computational energy burned during every orchestration cycle has become a critical engineering concern~\cite{devries2023growing}, with researchers emphasizing that runtime efficiency and hardware transparency must become first-class architectural priorities~\cite{HANKENDI2025113705}.
Despite their rapid adoption, most agentic AI frameworks are not yet mature enough for robust production use~\cite{11316910}. Practitioners deploy these frameworks without evaluating their inherent Technical Debt (TD). These design shortcuts inflate future maintenance costs. A highly measurable subset of this burden is Self-Admitted Technical Debt (SATD): shortcuts and known deficiencies that developers explicitly acknowledge in source code comments \cite{alves2014towards}. While TD is a well-established driver of software maintenance costs~\cite{cunningham1992wycash,alves2014towards}, its environmental impact in AI systems remains unexplored. Existing benchmarks primarily evaluate task completion rather than efficiency, failing to distinguish between highly optimized code and resource-intensive executions~\cite{NEURIPS2024_EffiBench}. As a result, the impact of technical debt on the runtime energy consumption of agentic frameworks is often ignored. Our study will investigate whether accumulated technical debt correlates with increased hardware-level energy overhead. If there is a strong correlation, SATD could serve as an accessible early-warning proxy for energy efficiency. Requiring only source code analysis, this proxy would allow developers to guide framework selection and prioritize energy-aware refactoring prior to any task execution. To automate SATD extraction across large codebases, we will adopt the established LLM-based approach proposed by Sheikhaei et al.~\cite{sheikhaei2024empirical}, which outperforms traditional baselines. We will conduct this investigation through a controlled empirical study across five open-source agentic frameworks drawn from a recent comparative analysis~\cite{11316910}. 
The remainder covers related work (\textbf{Section \ref{sec:related_works}}), empirical study design (\textbf{Section \ref{sec:study}}), potential threats to validity (\textbf{Section \ref{sec:threats}}), and expected contributions to energy-aware AI development (\textbf{Section \ref{sec:conclusion}}).

\section{Background and Related Work}
\label{sec:related_works}
\subsection{Self-Admitted Technical Debt in ML Systems}

Technical debt (TD) refers to long-term costs from short-term implementation shortcuts~\cite{cunningham1992wycash}, and Self-Admitted Technical Debt (SATD) serves as a useful indicator for studying TD, where developers explicitly acknowledge suboptimal implementations through code comments, issue trackers, or commit messages~\cite{alves2014towards}. While SATD in traditional software is well-studied~\cite{bavota2016large}, ML systems differs substantially. OBrien et al. mined 68,820 SATD comments from 2,641 ML repositories, proposing 23 ML-specific SATD types with Data Dependency debts being most prevalent~\cite{obrien202223}. Bhatia et al. found ML projects carry twice the SATD percentage of non-ML projects, and also identified two new ML-specific categories: Configuration Debt and Inadequate Testing Debt~\cite{bhatia2023empirical}. Liu et al. found Design Debt dominant in seven DL frameworks, alongside a newly identified Algorithm Debt~\cite{liu2020using}. Despite this body of work, no study has examined SATD in agentic AI systems, where LLM-based agents autonomously plan, use tools, and execute multi-step tasks, introducing unique concerns like agent orchestration, tool integration, and memory management that existing taxonomies do not cover. Based on established patterns across traditional and ML systems, Design Debt~\cite{maldonado2015detecting,liu2020using}, Configuration Debt~\cite{bhatia2023empirical}, and Data Dependency Debt~\cite{obrien202223} are expected to be the dominant SATD categories in agentic frameworks, given their architectural complexity, configurable LLM pipelines, and continuous data flow across agents, tools, and memory modules. We also anticipate that agentic systems will introduce novel, uncaptured challenges like multi-agent coordination, tool integration \cite{Abou_Ali2025} and prompt dependencies~\cite{aljohani2025promptdebt}.

\subsection{Sustainability and Energy Consumption in Agentic Workflows}
The rapid growth of LLMs has highlighted AI sustainability; generative models consume 29.3 TWh annually, yet their environmental impact remains underexplored~\cite{energycosts2025}. This gap is critical for agentic frameworks, which orchestrate LLMs for multi-step tasks~\cite{Abou_Ali2025}, compounding energy costs across every tool call and reasoning step~\cite{bao2026energyfootprintllmbased}. Recent studies state that the architecture of the framework directly influences energy consumption~\cite{tripathy2025swenergy}, positioning TD and energy efficiency as two deeply intertwined dimensions of the quality of the framework. Despite this, existing evaluations like AgentBench~\cite{liu2024agentbench} assess performance but omit energy metrics. Because the continuous orchestration of reasoning and tool execution in agentic loops significantly compounds energy overhead~\cite{tripathy2025swenergy}, fine-grained energy profiling is essential. Tools like \textit{pyRAPL}~\cite{pyrapl_docs} for CPU measurement and \textit{pynvml}~\cite{nvidia_ml_py} for GPU power monitoring are required to accurately capture the energy footprint~\cite{empiricalstudyDLmodels2024} of agentic task executions. 

While technical debt has been shown to increase maintenance costs in ML systems~\cite{bhatia2023empirical,liu2020using} and agentic frameworks compound energy across every orchestration cycle~\cite{bao2026energyfootprintllmbased,tripathy2025swenergy}, no study has examined whether SATD drives this energy overhead which is a gap that directly motivates \textbf{RQ3} (Section \ref{sec:RQ}). 
If a strong correlation exists, SATD can serve as a practical proxy for energy efficiency, integrating sustainability into standard code quality assessments without requiring expensive runtime profiling.

\section{Empirical Study Design}
\label{sec:study}
\subsection{Goal}
\textbf{Analyze} the technical debt and energy consumption induced by agentic AI frameworks
\textbf{for the purpose of} investigating the correlation between SATD and runtime energy impact
\textbf{with respect to} structural code quality and software sustainability
\textbf{from the viewpoint of} software engineering and AI researchers
\textbf{in the context of} standardized task execution.
\subsection{Research Questions}\label{sec:RQ}
To empirically investigate the relationship between software maintainability and energy cost, we propose a confirmatory study driven by three core research questions: \\
\textbf{RQ1 (SATD)}: How prevalent is SATD in open-source agentic AI frameworks, and what categories of TD are observed in these frameworks?\\
\textbf{RQ2 (Energy)}: How does the choice of agentic framework influence runtime energy consumption during standardised task execution?\\
\textbf{RQ3 (Correlation)}: To what extent do the density and categorical distribution of SATD in agentic framework source code relate to runtime energy consumption?

\subsubsection{Hypotheses}
RQ1 is descriptive and explores the landscape of Self-Admitted Technical Debt (SATD) in agentic AI frameworks. Hence, no formal hypotheses are defined.\\
\textbf{H1 (derived from RQ2)}: There is a measurable variance in runtime energy consumption across agentic AI frameworks, with framework architecture being a contributing factor.\\
\textbf{H2a (derived from RQ3)}: There is a statistically significant correlation between a framework's overall SATD density and its runtime energy consumption. \\
\textbf{H2b (derived from RQ3)}: The dominant category of SATD present in an agentic framework significantly influences its runtime energy consumption.
\subsection{Study Overview}
\begin{figure*}[!t]
    \centering
    \includegraphics[width=\textwidth]{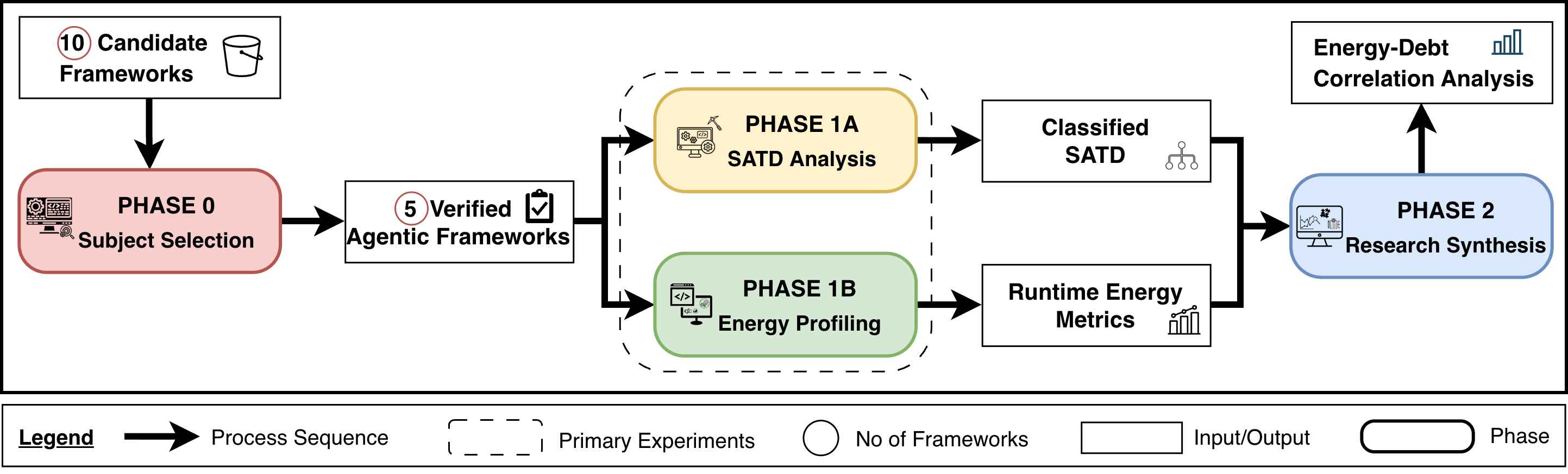}
    \caption{High-level Overview of the Study Design.}
    \label{fig:study_design_overview}
\end{figure*}
The three research questions demand two fundamentally different types of measurement, namely the static assessment of code quality and the dynamic profiling of runtime behaviour, which must be kept independent to avoid interference. Hence, we designed a multi-phase experimental workflow (Figure~\ref{fig:study_design_overview}) comprising \textbf{Phase 0} (\textit{Subject Selection}), \textbf{Phase 1A} (\textit{SATD Analysis}), \textbf{Phase 1B} (\textit{Energy Profiling}), and \textbf{Phase 2} (\textit{Research Synthesis}). Phases 1A and 1B execute independently in parallel, addressing \textbf{RQ1} and \textbf{RQ2} respectively, before converging in Phase 2 for the correlation analyses addressing \textbf{RQ3}.
\subsection{Phase 0: Subject Selection}
\label{sec:phase0}
To arrive at the \textbf{five} frameworks evaluated in this study, we first established a broader initial pool. The selection of agentic AI frameworks is a foundational methodological decision that determines the validity of all subsequent analyses. Rather than choosing by convenience, we apply a two-level protocol, \textit{evidence-based filtering} and \textit{baseline functional verification}, to \textbf{ten} candidates identified from recent literature~\cite{11316910}, ensuring subjects are \textit{representative}, \textit{instrumentable}, and \textit{diverse}~\cite{wohlin2012experimentation}. This is critical for \textbf{RQ3}; a homogeneous pool would suppress variance making correlation detection impossible. All artifacts and data related to this selection process are available in our replication package\footnote{https://doi.org/10.5281/zenodo.19550875} to enable open verification.
\subsubsection{Level 1 Evidence Based Selection}
We evaluate \textbf{ten} candidate frameworks against three exclusion criteria: (1) \emph{Observability}, requiring internal hooks for per-task energy and token logging; (2) \emph{Local Deployability}, requiring execution on local hardware to enable process-level energy measurement via RAPL/NVML (precluding cloud-only architectures); and (3) \emph{Architectural Distinctiveness}, where redundant frameworks with overlapping patterns or corporate backing are excluded in favor of higher-adoption alternatives. As shown in Table~\ref{tab:subject_selection}, applying these criteria narrows the pool to \textbf{six} frameworks.
\subsubsection{Level 2 Baseline Functional Verification}
The selected candidates will undergo local deployment and functional verification. A framework will be considered verified if it (i) \emph{installs without fatal dependency conflicts}; (ii) \emph{instantiates its core agent class in memory}; and (iii) \emph{executes a logical flow up to the external API authentication layer}. To isolate orchestration from inference, a successful outbound API dispatch was treated as a verified outcome. As shown in Table~\ref{tab:subject_selection}, AutoGPT was excluded at this stage due to its mandatory Docker-based agent runtime. Unlike task-level containerization, AutoGPT’s architecture precludes the step-level token logging and process-level energy measurement required by our protocol~\cite{pereira2017} without violating our no-modification constraint. Consequently, following is the final experimental pool ($n=5$) that captures the diverse architectural paradigms established in~\cite{liu2025agentdesignpattern,11316910}: \textbf{AutoGen} (multi-agent), \textbf{LangGraph} (graph-based), \textbf{Agno} (hierarchical), \textbf{Google ADK} (event-driven), and \textbf{SmolAgents} (lightweight). This wide range of architectures will maximize differences in both SATD density and energy use. Consequently, this variance ensures that the confirmatory correlation analyses for \textbf{RQ3} can detect meaningful relationships, even with a small sample size. To address temporal validity, each framework will be analysed across two version-pinned snapshots: \emph{Version 1} is pinned to April $11, 2026$, with exact commit hashes documented in the replication package; \emph{Version 2} will be pinned to the next stable release of each framework following a minimum interval of three to four months from \emph{Version 1}. These frameworks undergo Phase 1A (\textit{Static Analysis}) and Phase 1B (\textit{Dynamic Analysis}) independently on each snapshot to ensure they operate on an identical, frozen codebase.

\begin{table}[t]
\caption{Phase 0 - Level 1 and Level 2: Framework selection and verification decisions}
\label{tab:subject_selection}
\centering
\renewcommand{\arraystretch}{1.25}
\begin{tabular}{@{} l l l @{}}
    \textbf{Framework} & \textbf{Decision} & \textbf{Primary Rationale} \\ \hline
    LangGraph~\cite{01_langchain_langgraph}   & Included & Graph-state machine; partial observability via LangSmith\\ \hline
    AutoGen~\cite{02_microsoft_autogen}     & Included & Multi-agent conversation; strong telemetry support\\ \hline
    SmolAgents~\cite{03_huggingface_smolagents}  & Included & Code-first minimal execution; lightweight instrumentation\\ \hline
    Agno~\cite{04_agno_agi}        & Included & Function-calling wrapper; LLM-native, minimal overhead\\ \hline
    Google ADK~\cite{05_google_adk_python}  & Included & Hierarchical sub-agent; modular tracing and policy hooks\\ \hline
    CrewAI~\cite{06_crewaiinc_crewai}      & \textit{Excluded} (L1) & Observability: insufficient internal tracing\\ \hline
    MetaGPT~\cite{07_foundationagents_metagpt}     & \textit{Excluded} (L1) & Observability: abstraction obscures state transitions \\ \hline
    AgentGPT~\cite{08_reworkd_agentgpt}    & \textit{Excluded} (L1) & Deployability: cloud-first, prevents local energy measurement\\ \hline
    Semantic Kernel~\cite{09_microsoft_semantickernel} & \textit{Excluded} (L1) & Redundancy: overlaps with AutoGen (Microsoft)\\ \hline
    AutoGPT~\cite{10_significantgravitas_autogpt}     & \textit{Excluded} (L2) & Deployability: mandatory Docker setup limits instrumentation\\ \hline
\end{tabular}
\end{table}

\subsection{Phase 1A: SATD Analysis}
\label{sec:phase1A}
\begin{figure}[!h]
    \centering
    \includegraphics[width=\columnwidth]{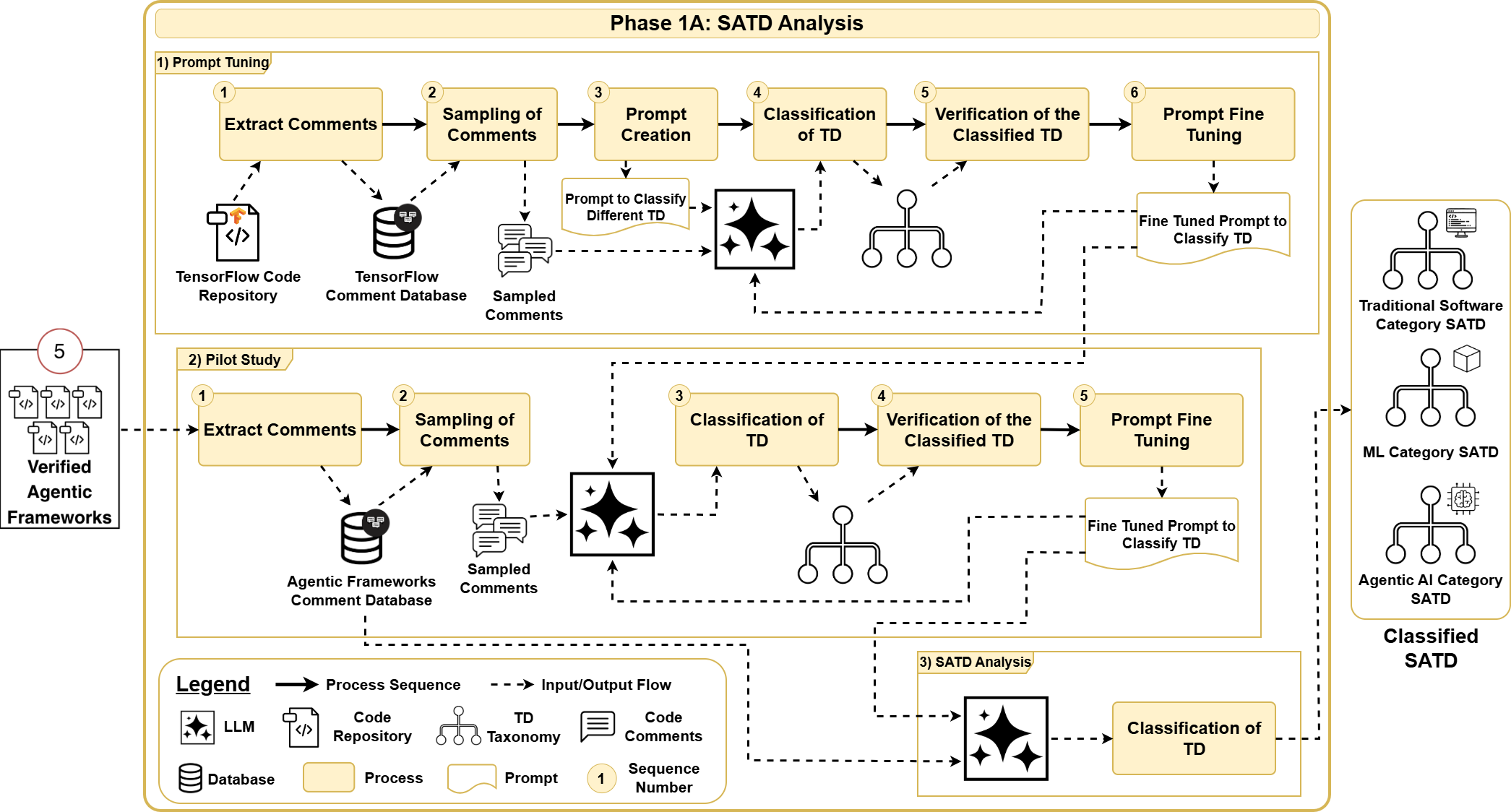}
    \caption{Phase 1A : SATD Analysis}
    \label{fig:Phase1A}
\end{figure}
Figure \ref{fig:Phase1A} presents the three-level SATD analysis workflow.The comment processing follows a two-step pipeline: (i) automated Python-based comment extraction from framework source repositories, requiring no framework execution, followed by (ii) LLM-based categorization using a fine-tuned prompt. Prior to LLM classification, comments undergo a four-step filtering pipeline adapted from Maldonado and Shihab~\cite{maldonado2015detecting}: (1) \textit{extract all source code comments}; (2) \textit{remove commented-out source code}; (3) \textit{remove license and header comments unless they contain task-reserved keywords (e.g., \emph{TODO}, \emph{FIXME)}}; and (4) \textit{merge consecutive single-line comments to preserve semantic context}.

\subsubsection{Level 1 Prompt Tuning}
We will use the TensorFlow repository as the calibration baseline~\cite{liu2020using} and adopt Bhatia et al.'s taxonomy~\cite{bhatia2023empirical} to categorize debt across traditional and ML-enabled systems. As shown in the \textit{\textbf{Prompt Tuning}} section of Figure \ref{fig:Phase1A}, all source code comments will be extracted into a TensorFlow Comment Database, and a representative sample drawn to keep manual verification manageable. An initial prompt is built on the TD Taxonomy, providing comment text plus surrounding code context~\cite{sheikhaei2024empirical}, and applied to the highest-ranked Arena leaderboard model~\cite{lmarena2026}. To establish a principled stopping criterion, we will evaluate each prompt iteration on a held-out sample across multiple LLMs (temperature=$0.0$, top-p=$0.1$) and measure inter-model agreement using Fleiss's Kappa. We will iteratively revise the prompt until theoretical saturation ($\kappa \ge 0.80$) is achieved. To ensure full reproducibility, all prompt versions, iteration logs, and per-model outputs will be documented in the replication package.

\subsubsection{Level 2 Pilot Study}
We then extract comments from the five frameworks (Table~\ref{tab:subject_selection}) into the Agentic Systems Comment Database. We will sample comments from this database, and will validate LLM classifications against the same $\kappa \ge 0.80$ threshold from \textit{Level 1}. Final prompt refinements will be done here before the main experiment.

\subsubsection{Level 3 SATD Analysis} 
Using the prompt refined through the pilot study, we apply it at scale to the complete Agentic Frameworks Comment Database to systematically identify and classify SATD instances. Each comment is provided to the LLM alongside its surrounding code context to improve classification accuracy. Rather than collapsing all debt into a single category, we adopt a three-dimensional classification scheme. The first dimension captures \textit{Traditional Software TD}, consistent with established SATD taxonomies~\cite{alves2014towards}. The second targets \textit{Machine Learning TD}, following the taxonomy proposed by \cite{bhatia2023empirical}. The third, \textit{Agentic TD}, may emerge given the nature of the systems under study, arising from challenges specific to agentic AI such as coordinating multiple agents and managing prompt dependencies~\cite{aljohani2025promptdebt}. The output of this phase is a labeled dataset of SATD instances drawn from real-world agentic framework repositories. To quantify the identified debt, we classify instances into nominal categories ($C_{type}$) and record their absolute count ($N_{SATD}$). To normalize comparisons across framework sizes, we calculate SATD density as the number of SATD instances per thousand lines of code (KLOC). To ensure consistency across the study, both comment extraction and KLOC calculations are restricted exclusively to Python source files. To validate LLM classifications, $384$ comments will be randomly sampled from the full output using Cochran's formula ($95$\% confidence, ±5\% margin of error). Two reviewers will independently label the sample; Cohen's Kappa will be reported for inter-rater agreement, with disagreements resolved by a second independent reviewer. The complete SATD analysis pipeline will be executed independently on both version-pinned snapshots, providing a temporal robustness check for \textbf{RQ1} prevalence findings and increasing the total pool of SATD instances across all categories.
\subsection{Phase 1B: Energy Profiling}
\label{sec:phase1B}
\begin{figure}[!h]
    \centering
    \includegraphics[width=\columnwidth]{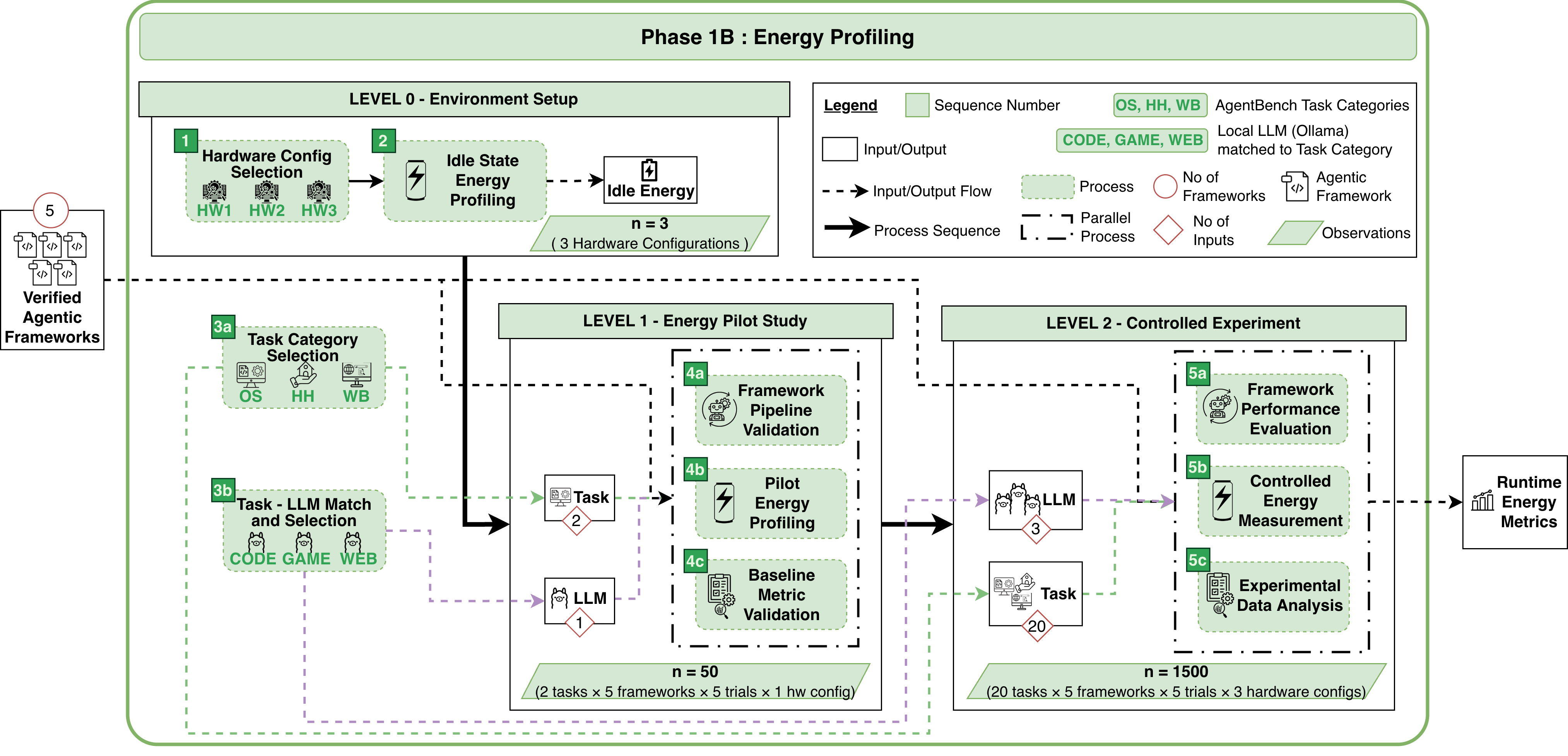}
    \caption{Phase 1B : Energy Profiling and Analysis}
    \label{fig:Phase1B}
\end{figure}

Executed in parallel with Phase 1A, \textbf{Phase 1B} will conduct the dynamic analysis to collect runtime energy consumption data across all \textbf{five} verified agentic frameworks under controlled and reproducible conditions. The focus will be on the orchestration layer (planning, memory, and tool dispatch); a domain-specific LLM will be held constant per task category to isolate framework efficiency from inference variation. The phase is structured into three sequential levels as illustrated in Figure \ref{fig:Phase1B}. Energy measurement will simultaneously employ \textit{pyRAPL}~\cite{pyrapl_docs} and \textit{pynvml}~\cite{nvidia_ml_py} to capture CPU and GPU energy consumption continuously throughout each task execution. Hardware-level measurement will be adopted over software estimation following the empirical methodology established by Pereira et al.~\cite{pereira2017}.

\subsubsection{Level 0 Environment Setup}
To ensure external validity, three distinct machines varying in CPU generation and GPU model will be provisioned : \emph{HW1} (Intel Xeon w3-2435, RTX A2000, 32/16 GB RAM/VRAM), \emph{HW2} (Intel i9-14900, RTX 4000 Ada, 64/20 GB RAM/VRAM), and \emph{HW3} (AMD Ryzen Threadripper PRO 5955WX, RTX A6000, 128/48 GB RAM/VRAM). To minimize energy variance, all non-essential background processes, network interfaces, and display services will be disabled. Before task execution, a 60-second idle energy reading will be captured per machine via \textit{pyRAPL} and \textit{pynvml}. By subtracting these values from subsequent measurements, we isolate the net task energy attributable to framework execution.
\subsubsection{Level 1 Energy Pilot Study}
To validate the measurement pipeline before the controlled experiment~\cite{wohlin2012experimentation}, Level 1 will serve as a pre-registered pilot study on the host operating system without Docker containerization, building on the environment configured in Level 0 (\textit{\textbf{Steps 1–2}}). We select OS Interaction (OS), Householding (HH), and Web Browsing (WB) from AgentBench~\cite{liu2024agentbench} as task categories (\textit{\textbf{Step 3a}}), with Level 1 focusing on two OS tasks ($n=2$) for measurement stability. Each category uses a domain-appropriate local LLM via Ollama~\cite{ollama}, with OS tasks using bash-tuned models and HH/WB using reasoning-optimized models (\textit{\textbf{Step 3b}}). Temperature will be fixed at $0.0$, seed at $42$, with identical system prompts. \textit{\textbf{Step 4}} verifies end-to-end pipeline execution per framework; two warm-up runs eliminate cold-start bias, with memory reset between trials. Upon successful pipeline verification, five measured trials per task-framework pairing will be conducted on \emph{HW1}, generating \textbf{50} total baseline observations. All runtime metrics (CPU/GPU energy, execution time, LLM call counts, and token throughput) must be captured successfully; any failure triggers a reconfiguration before advancing to Level 2. \textit{pyRAPL}, \textit{pynvml} and Ollama will be pinned to their latest stable releases at the start of Phase 1B, with exact versions explicitly documented prior to execution.
\subsubsection{Level 2 Controlled Experiment}
Upon successful validation from Level 1, the full experiment will execute within isolated Docker containers managed to prevent shared state dependencies. A wrapper script will enforce a strict execution sequence: container initialisation, idle baseline capture, task dispatch, parallel energy logging, result recording, and container teardown. \textit{\textbf{Step 5}} will dispatch all 20 AgentBench tasks (\textit{\textbf{Step 3a}} and \textit{\textbf{3b}}) across five frameworks, five trials, and three hardwares configurations, yielding \textbf{1,500} observations. Task outcomes will be recorded as binary pass/fail via AgentBench's automated grader. To isolate framework's orchestration overhead, the Ollama inference process will be tracked separately by PID; framework's induced energy will be calculated as the total measured energy minus the Ollama process energy. From this, two primary metrics will be computed via \textit{pyRAPL} and \textit{pynvml}: Total Energy per Task ($E_r$, in Joules) and Mean Energy per Framework ($\bar{E}_f$), which serves as the dependent variable for \textbf{H1, H2a, and H2b}. Secondary metrics, such as LLM call counts, energy per token, and run-to-run variance, will be tracked to isolate orchestration overhead. Statistical tests will be executed independently across all three hardware configurations; convergent results will confirm hardware-independent variance.  
\subsection{Phase 2: Research Synthesis}
\label{sec:phase2}
\textbf{Phase 2} examines the relationship between technical debt and energy consumption across agentic frameworks through two primary statistical, determining whether SATD derived from source code can serve as a practical proxy for predicting energy efficiency:

\textbf{Spearman's Rank Correlation for SATD Density and Energy Consumption:} To evaluate the relationship between SATD density and mean energy induced per framework without assuming a normal distribution, we will employ Spearman's Rank Correlation Coefficient ($\rho$). \textbf{Hypothesis H2a} will be tested at $\alpha = 0.05$ and will be supported if $p < 0.05$ and $\rho > 0$, confirming higher technical debt correlates with increased energy costs. Additionally, the effect size ($\rho^2$) will be reported to quantify the variance in energy consumption associated with SATD.

\textbf{Correlation for Distribution of SATD Categories and Energy Consumption:} 
To determine how the categorical distribution of SATD relates to energy consumption (Hypothesis H2b), we compute an independent Spearman's Rank Correlation ($\rho$) for each identified SATD category, correlating that category's density across the five frameworks with their corresponding mean energy consumption. Analysing categories independently prevents larger categories from dominating the analysis and masking the signal of less prevalent ones. Effect sizes ($\rho^2$) are reported alongside p-values for each category, as pre-specified. 

\section{Potential Threats to Validity}
\label{sec:threats}
\emph{\textbf{Construct validity:}} SATD detection will rely on keyword-based comment mining, which may miss implicit technical debt. The LLM will be provided with surrounding code context alongside comments~\cite{sheikhaei2024empirical}, enabling inference of debt lacking explicit keyword annotation. Confirmation bias in prompt tuning is mitigated via a pre-specified saturation protocol (Fleiss's $\kappa \geq 0.80$, temperature=$0.0$, top-p=$0.1$); a structured filtering pipeline and manual validation of a statistically derived sample ($384$ comments) further bound hallucination risk.
\emph{\textbf{Internal validity:}} Thinking token volume may inflate energy measurements independently of orchestration quality, potentially confounding the SATD–energy relationship. LLM call counts and input/output token counts will therefore be captured per trial, enabling token-normalised comparisons across frameworks.
\emph{\textbf{External validity:}} Findings are scoped to five open-source, Python-based frameworks paired with locally-served LLMs via Ollama, limiting generalisation to proprietary or cloud-scale deployments; replication across three hardware configurations partially addresses hardware-specific concerns. Temporal validity is bounded by two version-pinned snapshots; any cross-version inconsistencies will be explicitly reported as limitations.
\emph{\textbf{Conclusion validity:}} With five frameworks as the unit of analysis, statistical power will be inherently limited. We will therefore adopt a pre-specified minimum effect size of $\rho$ $\geq$ 0.50 following Cohen's large effect size convention~\cite{cohen1988statistical}, prioritising effect magnitude over p-value significance. For \textbf{H2a}, $\rho^2$ will be reported as a single effect size; for \textbf{H2b}, $\rho^2$ will be reported independently per SATD category to prevent dominant categories from masking less prevalent categories. Results must converge across \emph{HW1}, \emph{HW2}, \emph{HW3} for \textbf{H2a} and \textbf{H2b} to be considered supported.

\section{Conclusion}
\label{sec:conclusion}
As agentic AI frameworks become increasingly prevalent, understanding what drives their energy consumption is no longer optional. This study will systematically investigate whether Self-Admitted Technical Debt (SATD), framework architecture, and task execution are meaningful contributors to runtime energy overhead. Through a controlled, reproducible experimental design across five frameworks, twenty tasks, and three hardware configurations, we will establish an empirical foundation to answer \textbf{RQ1}, \textbf{RQ2}, and \textbf{RQ3}. The key findings will clarify whether SATD acts as a reliable, early-warning proxy for energy efficiency, or if inherent orchestration characteristics dominate the energy footprint. In doing so, this study will yield three core contributions: a reproducible methodology for measuring orchestration overhead, a baseline characterization of SATD in open-source agentic frameworks, and evidence-based guidance for engineering greener agentic workflows.


\bibliography{main}

\end{document}